\documentclass[aps, prd, superscriptaddress, preprintnumbers, floatfix, longbibliography, twocolumn, 10pt]{revtex4-2}
\usepackage{graphicx,float,wrapfig,subfigure}
\usepackage{amsfonts,amsmath,amssymb,amstext}
\usepackage{latexsym}
\usepackage{bm}
\usepackage{color}
\usepackage[normalem]{ulem}
\usepackage{MnSymbol}
\usepackage[colorlinks=true,linktocpage=true,linkcolor=blue,citecolor=blue]{hyperref}

\newcommand{\be}{\begin{equation}}
\newcommand{\ee}{\end{equation}}
\newcommand{\ba}{\begin{eqnarray}}
\newcommand{\ea}{\end{eqnarray}}

\definecolor{red}{rgb}{0.7,0,0}
\definecolor{green}{rgb}{0,0.5,0}

\begin{document}
 
 \title{Electrical conductivity of hot relativistic plasma in a strong magnetic field}
  
 \author{Ritesh Ghosh}
 \email{Ritesh.Ghosh@asu.edu}
 \affiliation{College of Integrative Sciences and Arts, Arizona State University, Mesa, Arizona 85212, USA}

 \author{Igor A. Shovkovy}
 \email{Igor.Shovkovy@asu.edu}
 \affiliation{College of Integrative Sciences and Arts, Arizona State University, Mesa, Arizona 85212, USA}
 \affiliation{Department of Physics, Arizona State University, Tempe, Arizona 85287, USA}
 
 \date{November 3, 2024}
 
 \begin{abstract}
We employ first-principles quantum field theoretical methods to investigate the longitudinal and transverse electrical conductivities of a strongly magnetized hot quantum electrodynamics (QED) plasma at the leading order in coupling. The analysis employs the fermion damping rate in the Landau-level representation, calculated with full kinematics and exact amplitudes of one-to-two and two-to-one QED processes. In the relativistic regime, both conductivities exhibit an approximate scaling behavior described by $\sigma_{\parallel,\perp} = T \tilde{\sigma}_{\parallel,\perp}$, where $\tilde{\sigma}_{\parallel,\perp}$ are functions of the dimensionless ratio $|eB|/T^2$ (with $T$ denoting temperature and $B$  magnetic field strength). We argue that the mechanisms for the transverse and longitudinal conductivities differ significantly, leading to a strong suppression of the former in comparison to the latter.
\end{abstract} 
 
 \maketitle
 
\section{Introduction}

Relativistic plasmas are common in cosmology, astrophysics, and heavy-ion collisions. In many cases, they coexists with strong magnetic fields \cite{Grasso:2000wj,Price:2006fi,Kharzeev:2015znc,Miransky:2015ava}. To understand the physical properties of such plasmas, one often needs to know their transport properties. One such defining property is electrical conductivity. It affects the decay (production) rate of magnetic fields, diffusion of charge fluctuations, and even the absorption (emission) of low-energy electromagnetic probes.

In the absence of background magnetic fields, the electrical conductivity of a relativistic QED plasma has been investigated by many authors \cite{Ahonen:1996nq,Baym:1997gq,Arnold:2000dr,Boyanovsky:2002te,Arnold:2003zc}. To some extent, it was also studied in weak magnetic fields \cite{vanErkelens:1984,Pike:2016aa}. However, there are no rigorous calculations of the electrical conductivity in strongly magnetized plasmas when the Landau-level quantization becomes important. Note, however, that some attempts have been undertaken in the context of quark-gluon plasma (QGP) using analytical methods \cite{Hattori:2016cnt,Hattori:2016lqx,Fukushima:2017lvb,Fukushima:2019ugr} and lattice calculations \cite{Aarts:2014nba,Brandt:2015aqk,Ding:2016hua,Astrakhantsev:2019zkr}. Here we address the problem from first principles in a weakly coupled plasma in a quantizing magnetic field. (We will also make an attempt to extrapolate the results to strongly coupled QGP, although it is formally outside the validity range of the approximations used.)

The common technique for calculating electrical conductivity is kinetic theory. Generally, however, it is not suitable for plasmas in quantizing magnetic fields when quantum states $\psi_{n}$ of charged particles are labeled by discrete Landau levels, $n=0,1,2,\ldots$, rather than continuous transverse momenta. In this case, one must use first-principles methods of quantum field theory to calculate transport properties. This is the approach we utilize in the current study to obtain electrical conductivity in a hot strongly magnetized QED plasma. We concentrate primarily on the electron-positron plasma in an ultra-relativistic regime with $\sqrt{|eB|}\gtrsim m_e$ and $T\gtrsim m_e$. However, the method is also valid at lower temperatures and weaker magnetic fields, provided 
$\sqrt{|eB|} \gtrsim \sqrt{\alpha} T$ \cite{Ghosh:2024owm}.

\section{Formalism}

Electrical conductivity is a measure of how easily current flows in response to an applied electric field.  At the microscopic level, it is determined by the damping rate (or the quasiparticle width) of charge carriers in the plasma. Generally, the smaller the damping rates of particles (which correlate with longer mean free paths), the greater the conductivity. In a strong magnetic field, this remains true for the longitudinal but not the transverse conductivity. Indeed, the magnetic field constrains the movement of charge carriers in perpendicular directions, resulting in a significant suppression of transport. As we explain below, it is the interactions responsible for particle damping that actually facilitate the flow of the transverse electrical current.

The fermion damping rate in a hot, strongly magnetized relativistic plasma has recently been calculated from  first principles in a gauge theory at the leading order in coupling \cite{Ghosh:2024hbf}. The resulting rate $\Gamma_{n}(k_z)$ depends on both the Landau-level index $n$ and the longitudinal momentum $k_z$.  In the presence of a sufficiently strong magnetic field, $\Gamma_{n}(k_z)$ is determined by the one-to-two and two-to-one processes: $\psi_{n}\to \psi_{n^\prime}+\gamma$, $\psi_{n}+\gamma \to \psi_{n^\prime}$, and $\psi_{n}+\bar{\psi}_{n^\prime}\to \gamma$, where $\gamma$ is a photon. In the absence of a magnetic field, these processes are forbidden due to the energy-momentum conservation. Consequently, in the weak field limit, the subleading two-to-two processes, $\psi_{n}+\gamma\to \psi_{n^\prime}+\gamma$ and $\psi_{n}+\bar{\psi}_{n^\prime}\to \gamma+\gamma$, must dominate. Roughly estimated, the magnitude of the leading-order contributions is $\alpha |eB|/T$, while the subleading processes are of the order $\alpha^2 T$, where $\alpha=1/137$ is the fine structure constant. In the rest of this study, therefore, we will assume that $|eB|/T^2\gg \alpha$.
 
To calculate the electrical conductivity tensor $\sigma_{ij}$, we use the Kubo linear-response theory. It relates the corresponding transport characteristics to the imaginary part of the polarization tensor, i.e.,  
\begin{equation}
\sigma_{ij}  = \lim_{\Omega\to 0}\frac{\mbox{Im}\,\Pi_{ij}(\Omega+i0 ;\bm{0})}{\Omega}.
\label{sigma-tensor}
\end{equation}
At one-loop order, the momentum-space expression for the polarization tensor is given by
\begin{equation}
\Pi_{ij}(\Omega;\bm{0}) =  4\pi \alpha \sumint_{k}  
\mbox{tr} \left[ \gamma^i \bar{G}(i\omega_k ;\bm{k})  
\gamma^j \bar{G}(i\omega_k-\Omega; \bm{k})\right],
\label{Pi-tensor}
\end{equation}
where $\bar{G}(i\omega_k ;\bm{k})$ is the Fourier transform of the translation invariant part of the fermion propagator \cite{Miransky:2015ava},
$\omega_k=\pi T (2k+1)$ is the fermion Matsubara frequency, and $\sumint_{k} \equiv T \sum_{k=-\infty}^{\infty} \int d^3 \bm{k}/(2\pi)^3$. 

Note that we omitted vertex corrections in Eq.~(\ref{Pi-tensor}). While this might raise concerns regarding gauge invariance, it has been demonstrated in Ref.~\cite{Hattori:2016cnt} that vertex corrections give only subleading contributions to conductivity in strong magnetic fields. This observation is also consistent with findings regarding plasmas without magnetic fields \cite{Aarts:2002tn}.

It is convenient to use the following spectral representation for the fermion propagator:
\begin{equation}
\bar{G}(i\omega_k ;\bm{k})  = \int_{-\infty}^{\infty} \frac{dk_{0} A_{\bm{k}} (k_0) }{i\omega_k-k_{0}+\mu}.
\label{prop-spectral-fun}
\end{equation}
Assuming that the electrical charge chemical potential $\mu$ is zero, the off-diagonal components of conductivity vanish. 
The remaining diagonal components are given by the following expression \cite{Gorbar:2013dha}
\begin{equation}
\sigma_{ii} = - \frac{\alpha}{8\pi T} \int \int \frac{d k_{0} d^3 \bm{k}}{\cosh^2\frac{k_{0}}{2T}}
\mbox{tr} \left[ \gamma^i A_{\bm{k}} (k_0) \gamma^i A_{\bm{k}} (k_0) \right].
\label{conductivity-ii}
\end{equation} 
In this representation, the spectral density $A_{\bm{k}} (k_0)$ contains all essential information about the damping rate of charge carriers  $\Gamma_{n}(k_z)$. The explicit expression of $A_{\bm{k}} (k_0)$ reads
\begin{eqnarray}
A_{\bm{k}} (k_0) &=& \frac{ie^{-k_\perp^2\ell^2}}{\pi}\sum_{\lambda=\pm}\sum_{n=0}^{\infty} 
 \frac{(-1)^n}{E_{n}}\Big\{
 \left[E_{n} \gamma^{0} 
 -\lambda  k_{z}\gamma^3+\lambda  m_e \right]
 \nonumber\\
&\times&
\left[{\cal P}_{+}L_n\left(2 k_\perp^2\ell^2\right)
-{\cal P}_{-}L_{n-1}\left(2 k_\perp^2\ell^2\right)\right] \nonumber\\
&+&2\lambda  (\bm{k}_\perp\cdot\bm{\gamma}_\perp) L_{n-1}^1\left(2 k_\perp^2 \ell^2\right)
 \Big\}\frac{\Gamma_{n}}{\left(k_{0} -\lambda E_{n} \right)^2+\Gamma_{n}^2} ,
 \label{spectral-density}
\end{eqnarray}
where $\ell = 1/\sqrt{|eB|}$ is the magnetic length, $E_{n}=\sqrt{2n|eB|+m_e^2+k_z^2}$ is the Landau-level energy, ${\cal P}_{\pm}=(1\mp i \gamma^1\gamma^2)/2$ are the spin projectors, and $L_n^{\alpha}\left(z\right)$ are the generalized Laguerre polynomials \cite{1980tisp.book.....G}. To simplify the notation, we suppressed the explicit dependence of $E_{n}$ and $\Gamma_{n}$ on the longitudinal momentum $k_z$. 

\begin{widetext}
Substituting spectral density (\ref{spectral-density}) into Eq.~(\ref{conductivity-ii}), and integrating over $\bm{k}_\perp$, we derive the final expressions for the transverse and longitudinal conductivities, i.e.,
\begin{eqnarray}
\sigma_{\perp} &=&
 \frac{\alpha}{\pi^2\ell^2 T}
\sum_{n=0}^{\infty}
\int_{-\infty}^{\infty} \int_{-\infty}^{\infty} \frac{d k_{0} d k_{z} }{\cosh^2\frac{k_{0}}{2T}}
 \frac{\Gamma_{n+1} \Gamma_{n}
 \left[ \left(k_{0}^2+E_n^2+\Gamma_n^2\right)\left(k_{0}^2+E_{n+1}^2+\Gamma_{n+1}^2\right)  -4k_{0}^2\left(k_z^2 +m_e^2\right)\right]}
{\left[\left(E_{n}^2+\Gamma_{n}^2 -k_{0}^2\right)^2+4k_{0}^2 \Gamma_{n}^2 \right]
\left[\left(E_{n+1}^2+\Gamma_{n+1}^2 -k_{0}^2\right)^2+4k_{0}^2 \Gamma_{n+1}^2 \right]},
\label{conductivity-11}
\\
\sigma_{\parallel} &=& \frac{\alpha}{2\pi^2\ell^2 T}
 \sum_{n=0}^{\infty}\beta_n
\int_{-\infty}^{\infty} \int_{-\infty}^{\infty} \frac{d k_{0} d k_{z} }{\cosh^2\frac{k_{0}}{2T}}
\frac{\Gamma_n^2\left[\left(E_n^2+\Gamma_n^2-k_{0}^2\right)^2 +4k_{0}^2\left(2k_z^2+\Gamma_n^2\right)\right]}
{\left[\left(E_{n}^2+\Gamma_{n}^2 -k_{0}^2\right)^2+4k_{0}^2 \Gamma_{n}^2 \right]^2} ,
\label{conductivity-33}
\end{eqnarray}
\end{widetext}
respectively. In the last expression, we used $\beta_n\equiv 2-\delta_{n,0}$. 

The longitudinal conductivity in Eq.~(\ref{conductivity-33}) is determined by a sum of contributions from individual Landau levels. The latter can be viewed formally as distinct species of particles. In the case of small damping rate, the integral over the energy is dominated by sharp peaks of the spectral density at $k_0=\pm E_{n}$. Then, it is easy to show that each level contributes proportionally to the inverse damping rate: $\sigma_{\parallel} = \sum_n \sigma_{\parallel,n}$, where $\sigma_{\parallel,n} \propto 1/\Gamma_{n}$.

In contrast, the nature of transverse conductivity differs substantially. The partial contributions in Eq.~(\ref{conductivity-11}) come from transitions between adjacent Landau levels. Assuming that the field is sufficiently strong and the damping rate is small, the contributions are proportional to the damping rates rather than their inverse values, i.e., $\sigma_{\perp} = \sum_n \sigma_{\perp,n}$, where $\sigma_{\perp,n} \propto \Gamma_{n}$. This reflects a unique conduction regime with the magnetic field confining charged particles in the transverse plane, while interactions providing a transport pathway via quantum transitions (``jumps") between Landau levels.

\section{Electrical conductivity in QED plasma}

The formal expressions for the transverse and longitudinal conductivities in Eqs.~(\ref{conductivity-11}) and (\ref{conductivity-33}) are only as good as the quality of input about the fermion damping rates $\Gamma_{n}(k_z)$. Here we rely on the recent state-of-the-art results for the Landau-level dependent damping rates obtained at the leading order in coupling in Ref.~\cite{Ghosh:2024hbf}. Because of weak coupling, the corresponding  approximation should be reliable in the case of QED plasma, assuming $|eB|/T^2\gg \alpha$. 

To calculate electrical conductivities, we begin by generating numerical data for the damping rates across a wide  range of magnetic fields (from $|eB|=225m_e^2$ to $|eB|=51200m_e^2$) and temperatures (from $T=20m_e$ to $T=80m_e$). Following the methodology outlined in Ref.~\cite{Ghosh:2024hbf}, we compute $\Gamma_{n}(k_z)$ for all Landau levels up to $n=50$. In our calculation, however, we account for all processes involving Landau-level states with indices up to $n_{\rm max} =100$. Such a large phase space is necessary to attain a good approximation for the longitudinal conductivity $\tilde{\sigma}_{\parallel}$ when $|eB|/T^2\gtrsim 0.1$, covering almost the whole range of validity of the leading-order approximation. Since the Landau-level sum in the transverse conductivity (\ref{conductivity-11}) converges slightly slower than in the longitudinal conductivity, one needs to include even more terms. The approximation with $50$ Landau levels works reasonably well for $\tilde{\sigma}_{\perp}$ only when $|eB|/T^2\gtrsim 0.3$ or so.

\begin{figure}[t]
	\centering
	\includegraphics[width=0.95\columnwidth]{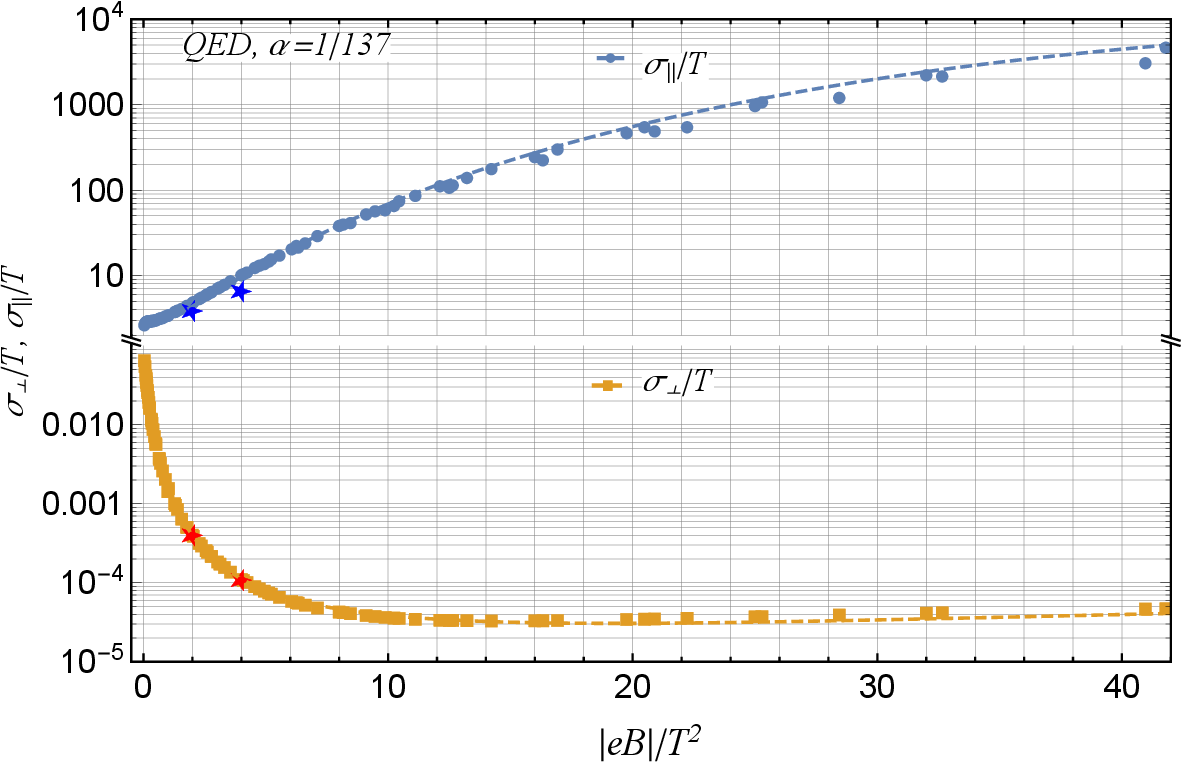}  
	\caption{The longitudinal and transverse conductivities in QED. The dashed lines represent the analytical fits in Eqs.~(\ref{signa11QEDfit}) and (\ref{signa33QEDfit}). Two extra points (labeled by stars) correspond to a low temperature, $T=2.5m_e$, and two relatively weak magnetic fields, $|eB| =12.5m_e^2 $ and $|eB|  =25m_e^2$.}
	\label{fig:QED}
\end{figure}

Using the numerical data for damping rates $\Gamma_{n}(k_z)$, we readily calculate the transverse and longitudinal conductivities, defined by Eqs.~(\ref{conductivity-11}) and (\ref{conductivity-33}), respectively. The results are summarized in Fig.~\ref{fig:QED}, where we show the dimensionless ratios $\tilde{\sigma}_{\perp}\equiv \sigma_{\perp}/T$ (orange) and $\tilde{\sigma}_{\parallel}\equiv \sigma_{\parallel}/T$ (blue) as functions of $|eB|/T^2$. (For the numerical data, see Supplemental Material \cite{DataCond:2024}.) 

As we see, despite a wide range of magnetic fields and temperatures, nearly all data fall on the same curve. This is not  surprising since the effects of a nonzero fermion mass are negligible in the ultrarelativistic regime with $\sqrt{|eB|/T}\gg m_e$ and $T\gg m_e$. In such a scaling regime, $|eB|/T^2$ is the only relevant dimensionless parameter that determines $\tilde{\sigma}_{\perp}$ and $\tilde{\sigma}_{\parallel}$. 

Of course, deviations from the universal scaling dependence in Fig.~\ref{fig:QED} become noticeable with decreasing of the temperature and/or the magnetic field. To demonstrate this, we included two such cases in Fig.~\ref{fig:QED}, labeled by stars. They represent conductivities at a relatively low temperature, $T=2.5m_e\approx 1.28~\mbox{MeV}$, and two rather weak magnetic fields, $|eB| =12.5m_e^2$ ($B\approx 5.52\times 10^{14}~\mbox{G}$) and $|eB| =25m_e^2$ ($B\approx 1.10\times 10^{15}~\mbox{G}$). We also verified that the deviations from the ultrarelativistic scaling become noticeable in the longitudinal conductivity $\tilde{\sigma}_{\parallel}$ even at large $\sqrt{|eB|/T^2}$ for moderately low temperatures, $T\lesssim 20 m_e$. 

In the scaling regime, the transverse and longitudinal conductivities can be fitted by the following Pad\'{e} approximants:
\begin{eqnarray}
\frac{\tilde{\sigma}_{\perp}^{\rm QED}}{ \tilde{\sigma}_{\perp,0}}&=& 
\frac{ 1+0.125 b^{3/2} + 0.00266 b^{7/2}}{1 + 6.5 b + 31.4 b^2 +  13.5 b^3}, 
\label{signa11QEDfit} \\
\frac{\tilde{\sigma}_{\parallel}^{\rm QED}}{ \tilde{\sigma}_{\parallel,0}} &=&  
\frac{ 1 + 0.09  b^2 + 0.0017 b^4}{1+0.11 b-0.003 b^{2} +9\times 10^{-5}  b^3} ,
\label{signa33QEDfit}
\end{eqnarray}
where $\tilde{\sigma}_{\perp,0}\approx 0.076 $, $ \tilde{\sigma}_{\parallel,0}\approx 2.4$, and $b=|eB|/T^2$. In the whole range with $0.1 \lesssim |eB|/T^2\lesssim 45$, the quality of the fit is about $4\%$ for $\tilde{\sigma}_{\perp}$ and $20\%$ for $\tilde{\sigma}_{\parallel}$.

In accordance with the very different underlying mechanisms mentioned earlier, the longitudinal conductivity $\tilde{\sigma}_{\parallel}$ is much higher than the transverse conductivity $\tilde{\sigma}_{\perp}$. In QED, their ratio reaches values of the order of $10^3$ already at $|eB|/T^2\simeq 1$ and grows to about $10^7$ at $|eB|/T^2\simeq 50$. This is consistent with a qualitative dependence on the coupling constant, provided $\sigma_{\perp} \propto \Gamma_{n} \propto \alpha$ and $\sigma_{\parallel} \propto 1/\Gamma_{n} \propto 1/\alpha$. 

The longitudinal conductivity grows monotonically with increasing $|eB|/T^2$. It might seem surprising that $\tilde{\sigma}_{\parallel}$ grows even in the region of small values of $|eB|/T^2$, where many Landau levels contribute. As the field grows, fewer Landau levels contribute to $\tilde{\sigma}_{\parallel}$. At the same time, their degeneracy increases linearly with the field and the damping rates $\Gamma_{n}(p_z)$ decrease (at small values of $p_z$). The latter two effects overcompensate the reduction in the number of Landau levels contributing to $\tilde{\sigma}_{\parallel}$. As a result, the longitudinal conductivity grows with the field.

In contrast, the transverse conductivity mostly decreases with increasing $|eB|/T^2$. To a large degree, this is the consequence of a very different transport mechanism facilitated by interactions. We find that, at small values of $|eB|/T^2$, the dominant contributions to $\tilde{\sigma}_{\perp}$ come from Landau levels with large indices, $n\sim T^2/|eB|$. Because of a slow convergence of the Landau-level sum in Eq.~(\ref{conductivity-11}), a much larger number of terms must be included to achieve the same precision as in the longitudinal conductivity. The transverse conductivity reaches a minimum around $b_{\rm min}\simeq 12$ and then starts to grow slowly. We find that it is dominated by transitions between the $0$th and $1$st Landau levels when $b\gtrsim b_{\rm min}$.

\section{Other magnetized plasmas}

A QED plasma composed of electrons and positrons is not the only type of magnetized plasma that plays a significant role in relativistic systems. Naturally, one might ask whether our calculations of electrical conductivity can be extended to other plasmas, such as QGP or the primordial plasma in the early Universe. Formally, this should be possible if the coupling constant is sufficiently weak.

Numerous theoretical models suggest that superstrong magnetic fields, on the order of $10^{23}~\mbox{G}$, were generated in the early Universe \cite{Grasso:2000wj}. These fields influenced the evolution of the primordial plasma, leaving observable imprints in the temperature and polarization anisotropies of the cosmic microwave background radiation. Additionally, they may provide an explanation for the origin of galactic and intergalactic magnetic fields in the present Universe \cite{Vachaspati:2020blt}.

High electrical conductivity is a key assumption in most cosmological studies, as it implies the conservation of magnetic flux and magnetic helicity \cite{Grasso:2000wj}. Conventionally, however, conductivity is estimated without considering the impact of the magnetic field. Our study demonstrates that strong magnetic fields can significantly affect charge transport. Since high temperatures ensure a weakly coupled regime, our first-principles approach for calculating electrical conductivity is likely applicable to the primordial plasma. Although this falls outside the scope of our current investigation, exploring the effects of anisotropic conductivity on early Universe physics in the presence of strong magnetic fields would be a valuable direction for future research.
 
It is also worth considering whether the same methods can be rigorously applied to calculate the electrical conductivity of QGP. Strongly magnetized QGP can be created in ultrarelativistic heavy-ion collisions, where electrical conductivity plays an important role in the underlying phenomenology \cite{Tuchin:2013apa,McLerran:2013hla,Gursoy:2014aka,Tuchin:2015oka}. It determines the relaxation time of the background magnetic field, which in turn affects various observable signatures, including those associated with the chiral magnetic effect and other related phenomena. (For reviews, see Refs.~\cite{Kharzeev:2015znc,Miransky:2015ava}.)

There have been many attempts in the literature to study the effect of a magnetic field on the electrical conductivity of QGP. The methods ranged from various versions of kinetic theory \cite{Fukushima:2017lvb,Ghosh:2019ubc,Dey:2019axu,Dey:2019vkn,Das:2019ppb} to field theory models \cite{Hattori:2016cnt,Satapathy:2021cjp,Peng:2023rjj,Bandyopadhyay:2023lvk}, holographic QCD \cite{Fukushima:2021got,Li:2018ufq}, and lattice QCD \cite{Buividovich:2010tn,Astrakhantsev:2019zkr}. Recently, also experimental constrains from the STAR collaboration have been reported in Ref.~\cite{STAR:2023jdd}.

In contrast to its QED counterpart, QGP is made of strongly interacting quarks and gluons, rendering perturbative techniques generally inapplicable. However, by invoking the principles of asymptotic freedom \cite{Gross:1973id,Politzer:1973fx}, one can argue that such a plasma becomes weakly interacting at extremely high temperatures \cite{Busza:2018rrf} and/or under superstrong magnetic fields \cite{Miransky:2002rp}. Then, one can use perturbative methods to illuminate transport properties of QGP. Formally, the conductivities are given by expressions similar to those in Eqs.~(\ref{conductivity-11}) and (\ref{conductivity-33}). However, to account for the quark electrical charges and colors, one should  replacing $\alpha$ with $\alpha N_c (q_f/e)^2$ and $|eB|$ with $|q_f B|$ ($f=u,d$). 
 
At the leading order in the coupling constant, $\alpha_s=g_s^2/(4\pi)$, the damping rates of quarks, $\Gamma_{n}^{f}(k_z)$, are determined by the following one-to-two and two-to-one processes: $\psi_{n}\to \psi_{n^\prime}+g$, $\psi_{n}+g \to \psi_{n^\prime}$,  and $\psi_{n}+\bar{\psi}_{n^\prime}\to g$, where $g$ represents a gluon. The corresponding rates are approximately of the order of $\alpha_s |eB|/T$ \cite{Ghosh:2024hbf}. However, this leading-order approximation is only valid if the running coupling constant remains small, $\alpha_s\ll 1$.

Building on these general, albeit simplistic, arguments, we can extend the perturbative field-theoretical calculations of electrical conductivity to a strongly magnetized QGP. Noting that the world average value of the running QCD coupling is $\alpha_s \left(M_Z^2\right)=0.1181 \pm 0.0011$ at the $Z$ boson mass scale \cite{ParticleDataGroup:2024cfk}, the temperature must satisfy $T\gtrsim M_Z$ to justify the validity of perturbative calculations in QCD. In contrast, the typical temperatures of QGP produced in ultrarelativistic heavy-ion collisions are around $T\simeq 300~\mbox{MeV}$, which are much lower, resulting in a running value of  $\alpha_s$ on the order of $1$.

Clearly, the leading-order approximation is inadequate for reliably estimating the electrical conductivity in a strongly interacting regime of QGP. Nevertheless, by formally extending our calculations to such a regime, we would find that the numerical results for $\sigma_\parallel$ are much larger than the results on the lattice \cite{Ghosh:2024owm}. This discrepancy is most likely due to large subleading corrections from two-to-two processes that cannot be ignored at strong coupling. Additionally, nonperturbative effects, such as the magnetic catalysis of chiral symmetry breaking \cite{Miransky:2015ava}, may play a role in a strongly magnetized QGP. In either case, it is clear that further research is necessary to explore all relevant effects at strong coupling. Nonetheless, it is intriguing to consider the possibility that the longitudinal conductivity is very large in a strong magnetic field. If it were the case, lattice calculations might be unable to study such transport.

Although our leading-order calculations cannot offer reliable quantitative predictions for heavy-ion physics, they do provide a deeper insight into  the electric charge transport in a strongly magnetized QGP at sufficiently high temperatures. As in QED, the underlying mechanisms responsible for the   transverse and longitudinal conductivities are very different in the presence of a strong magnetic field. While the former tends to be enhanced by interactions ($\sigma_{\perp,n} \propto \Gamma_{n}$), the latter is  suppressed  by them ($\sigma_{\parallel,n} \propto 1/\Gamma_{n}$). In principle, this can be verified by using lattice methods in various QCD-like theories with different coupling constants $\alpha_s$. According to our predictions, the ratio of the transverse to the longitudinal conductivity should scale roughly as $\alpha_s^2$. 

In connection with the electric conductivity of QGP, it is instructive to compare our results with earlier gauge theory studies \cite{Hattori:2016cnt,Hattori:2016lqx,Fukushima:2017lvb,Fukushima:2019ugr}, which bare some similarities to our approach.  The authors of Ref.~\cite{Hattori:2016cnt} employ a method similar to ours but focus only on the longitudinal transport and utilize a simpler approximation for the fermion damping rates. We verified that the lowest Landau-level approximation of Ref.~\cite{Hattori:2016lqx} is consistent with our narrow spectral-peak approximation for $\sigma_\parallel$ when expressed in terms of the damping rate. However, our analysis indicates that a naive lowest Landau-level approximation for the rate, which ignores transitions to other levels, is unreliable.

In Refs.~\cite{Fukushima:2017lvb,Fukushima:2019ugr}, the authors develop a special formulation of kinetic theory for longitudinal transport in a strong magnetic field. While the shape of their scaling function $\tilde{\sigma}_{\parallel}$ resemble ours at strong fields, it differs significantly in magnitude and detail. We believe their method for calculating $\tilde{\sigma}_{\parallel}$ should be formally equivalent to our narrow spectral-peak approximation. However, the use of an artificial choice of power-law basis functions for calculating matrix elements of the collision integral in Refs.~\cite{Fukushima:2017lvb,Fukushima:2019ugr}, might render the method numerically unreliable. Also, the kinetic theory method of Refs.~\cite{Fukushima:2017lvb,Fukushima:2019ugr} cannot be applied to transverse transport at all. 

\section{Summary}

In this study, we used first-principles quantum-field theoretical methods to calculate the electrical conductivity of strongly magnetized relativistic plasmas. Our calculation relies on the Kubo formula and utilizes leading-order results for the Landau-level dependent fermion damping rates $\Gamma_{n}(p_z)$. For calculating $\Gamma_{n}(p_z)$, we utilized the formalism developed recently in Ref.~\cite{Ghosh:2024hbf}. In a sufficiently strong magnetic field, the damping rates are dominated by the one-to-two and two-to-one QED processes: $\psi_{n}\to \psi_{n^\prime}+\gamma$, $\psi_{n}+\gamma \to \psi_{n^\prime}$,  and $\psi_{n}+\bar{\psi}_{n^\prime}\to \gamma$. The corresponding leading-order analysis should be valid when the magnetic field and temperature satisfy the inequality $|eB|/T^2\gg \alpha $. For smaller ratios, the approximation breaks down because the damping rates are dominated by the subleading two-to-two processes, $\psi_{n}+\gamma\to \psi_{n^\prime}+\gamma$ and $\psi_{n}+\bar{\psi}_{n^\prime}\to \gamma+\gamma$.  

For simplicity, here we concentrated primarily on the ultrarelativistic QED plasma, which is realized when $T\gg m_e$ and $\sqrt{|eB|}\gg m_e$. In this regime, the ratios of the longitudinal and transverse conductivities to the temperature, $\sigma_{\perp}/T$ and $\sigma_{\parallel}/T$, can be approximated by universal scaling functions $\tilde{\sigma}_{\perp}$ and $\tilde{\sigma}_{\parallel}$ that depend only on  $|eB|/T^2$. The corresponding numerical results are presented in Fig.~\ref{fig:QED}. 

As expected, electric charge transport reveals a high degree of anisotropy in a strong magnetic field. In QED, the  values of $\tilde{\sigma}_{\perp}$ are 2 to 7 orders of magnitudes smaller than $\tilde{\sigma}_{\parallel}$. More importantly, the underlying mechanisms for transport differ significantly in the transverse and longitudinal directions. Without interactions, the motion of charge carriers is unimpeded in the direction parallel to the magnetic field but highly restricted in the perpendicular directions. The effect of interactions is opposite. While particle scattering into other states suppresses longitudinal transport, transitions between different Landau-level orbitals facilitate transverse transport. Consequently, the longitudinal conductivity scales with the coupling constant as $1/\alpha$, but the transverse conductivity scales as $\alpha$, resulting in a huge difference between the two. 

We note that reliable results for the conductivity are crucial for realistic simulations of pulsar magnetospheres \cite{Li:2011zh}. Specifically, the inverse of $\tilde{\sigma}_{\parallel}$ quantifies the deviation from the force-free condition in the plasma, the magnitude of the parallel component of the electric field, and the current dissipation rate. These factors, in turn, influence the activity of the magnetosphere and the observational signatures of pulsars \cite{Kalapotharakos:2011vg,Kalapotharakos:2012dq}.

To extrapolate our leading-order calculations to the case of strongly magnetized QGP, it is essential to ensure that the QCD running coupling constant remains sufficiently weak. However, under conditions relevant to heavy-ion collisions, the coupling constant $\alpha_s$ is typically on the order of $1$. Clearly, at such large coupling, the leading-order approximation breaks down. In this regime, only non-perturbative techniques can provide reliable quantitative results. Nevertheless, our analysis may still offer valuable insights into the qualitative differences between longitudinal and transverse transport in strongly magnetized QGP, which could play an important role in heavy-ion phenomenology. 

Investigating charge transport of QGP remains an important task for future research. Indeed, in heavy-ion collisions, electrical conductivity plays a critical role in the evolution of plasma. It defines the trapping and diffusion scales of magnetic fields within the QGP and influences various charged particle correlations. The first experimental measurements of conductivity were reported by the STAR collaboration in Ref.~\cite{STAR:2023jdd}. A better understanding of electrical conductivity is crucial for providing reliable theoretical predictions, particularly in relation to detecting signatures of the chiral magnetic effect \cite{Kharzeev:2024zzm}.
\medskip

\begin{acknowledgments}
This research was funded in part by the U.S. National Science Foundation under Grant No.~PHY-2209470.
\end{acknowledgments}

\end{document}